\documentclass[prl,aps,twocolumn,superscriptaddress,floatfix,showpacs]{revtex4}

\newcommand{\bind}[1]{E_{b}^{\mathrm{#1}}}
\newcommand{\bexc}{\bind{exc}}
\newcommand{\bpol}{\bind{pol}}
\newcommand{\vE}{\mathbf{E}}
\newcommand{\vD}{\mathbf{D}}
\newcommand{\om}{\omega}
\newcommand{\omn}{\om_n}
\newcommand{\eps}{\epsilon}

\newcommand{\epsi}{\eps_{\infty}}
\newcommand{\epss}{\eps_{s}}
\newcommand{\tad}{\tau_{D}}
\newcommand{\tal}{\tau}
\newcommand{\taln}{\tau_n}

\newcommand{\epom}{\eps (\om)}
\newcommand{\aom}{\alpha (\om)}
\newcommand{\aomn}{\alpha (\omn)}
\newcommand{\alo}{\alpha_{0}}
\newcommand{\Alt}{\mathcal{A}(t)}
\newcommand{\xo}{x_{0}}
\newcommand{\xp}{x^{\prime}}
\newcommand{\tp}{t^{\prime}}

\newcommand{\wfxt}{\psi(x,t)}
\newcommand{\wfx}{\psi(x)}
\newcommand{\wfxp}{\psi(x^{\prime})}
\newcommand{\potxt}{\phi(x,t)}
\newcommand{\dpotxt}{\tilde{\phi}(x,t)}
\newcommand{\dpotxpt}{\tilde{\phi}(\xp,t)}
\newcommand{\potfxt}{\phi_{f}(x,t)}
\newcommand{\potf}{\phi_{f}}
\newcommand{\potaxt}{\phi_{a}(x,t)}
\newcommand{\potax}{\phi_{a}(x)}
\newcommand{\pota}{\phi_{a}}
\newcommand{\potixt}{\phi_{i}(x,t)}
\newcommand{\potsxt}{\phi_{s}(x,t)}
\newcommand{\potsko}{\phi_{s}(k,\om)}
\newcommand{\denxt}{\rho(x,t)}

\newcommand{\denxpt}{\rho(\xp,t)}
\newcommand{\den}{\bar{\rho}}
\newcommand{\denp}{\bar{\rho}\,^{\prime}}
\newcommand{\dden}{\tilde{\rho}}
\newcommand{\pot}{\bar{\phi}}
\newcommand{\potp}{\bar{\phi}\,^{\prime}}
\newcommand{\denn}{\rho_{n}}
\newcommand{\deno}{\rho_{0}}
\newcommand{\denm}{\rho_{m}}
\newcommand{\potn}{\phi_{n}}
\newcommand{\potnp}{\phi_{n}^{\prime}}
\newcommand{\poto}{\phi_{0}}
\newcommand{\lan}{\lambda_{n}}
\newcommand{\sumn}{\sum_{n \geq 1}}
\newcommand{\ksit}{\xi (t)}
\newcommand{\dxi}{\dot{\xi}}
\newcommand{\ant}{a_{n} (t)}
\newcommand{\an}{a_{n}}

\newcommand{\xrelt}{x-\ksit}
\newcommand{\ga}{\gamma}

\newcommand{\avg}[1]{\langle #1 \rangle}
\newcommand{\kB}{k_{B}}
\newcommand{\fE}{\mathcal{E}}
\newcommand{\aB}{a_{B}}
\newcommand{\est}{\epsilon^{*}}
\newcommand{\Ry}{\mathrm{Ry}}

\usepackage{graphicx}
\usepackage{amsmath}
\usepackage{latexsym}

\begin{document}

\title{One-dimensional semiconductor in a polar solvent: Solvation and low-frequency
dynamics of an excess charge carrier}

\author{Yu.~N.~Gartstein}
\affiliation{Department of Physics, The University of Texas at
Dallas, P. O. Box 830688, FO23, Richardson, Texas 75083, USA}
\author{G.~L.~Ussery}
\affiliation{Department of Physics, The University of Texas at
Dallas, P. O. Box 830688, FO23, Richardson, Texas 75083, USA}

\begin{abstract}
Due to solvation, excess charge carriers on 1$d$ semiconductor nanostructures immersed in polar solvents undergo self-localization into polaronic states. Using a \textit{simplified} theoretical model for small-diameter structures, we study low-frequency dynamical properties of resulting 1$d$ adiabatic polarons. The combined microscopic dynamics of the electronic charge density and the solvent leads to macroscopic Langevin dynamics of a polaron and to the appearance of local dielectric relaxation modes. Polaron mobility is evaluated as a function of system parameters. Numerical estimates indicate that the solvated carriers can have mobilities orders of magnitude lower than the intrinsic values.
\end{abstract}

\maketitle

\section{Introduction}

\begin{figure}
\includegraphics[scale=0.7]{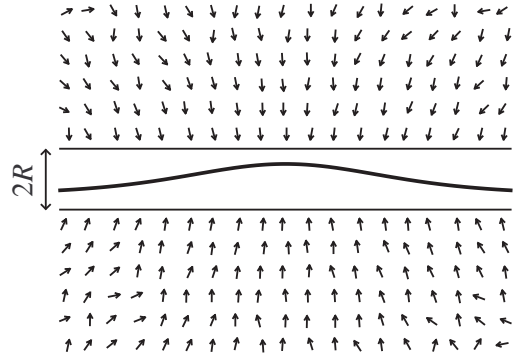}
\caption{\label{polimage}Schematically: Solvated excess electron charge density (thick line) along the 1$d$ semiconductor nanostructure which is surrounded by a pattern of the orientational polarization of solvent molecules.}
\end{figure}

One-dimensional (1$d$) semiconductor (SC) nanostructures like nanowires and nanotubes are fascinating objects offering a promise for various applications, including electronics, energy harvesting and sensors. The properties of excess charge carriers are fundamental for these applications. A customary picture used to describe charge carriers in nanowires \cite{NWelectronics} and nanotubes \cite{NTelectronics} is that of band electron states with very high, nearly ballistic \cite{dressbook1}, intrinsic mobilities.
Many applications, however, such as photoelectrochemical energy conversion \cite{kamat07,pyang05}, feature SC nanostructures in contact with polar liquid environments. We have recently pointed out \cite{YNGpol,polcylinder} that excess carriers on 1$d$ SCs immersed in the sluggish polar medium can drastically change their nature. Due to the long-range Coulomb interaction with the slow (orientational) polarization of the medium, a charge carrier (an electron or a hole) gets solvated forming a localized electronic state surrounded by a self-consistent dielectric polarization pattern as shown schematically in Fig.~\ref{polimage}. In physical parlance, self-trapped states formed by this mechanism are traditionally called polarons, extensively studied in 3$d$ ionic crystals and polar SCs \cite{polarons1,alexmott}. Another related 3$d$ notion is that of the solvated electron in liquids \cite{ferra91,CTbook,nitzan}.

The formation of 1$d$ polarons is expected to have an impact on various processes such as electron-hole separation from bound excitonic states \cite{YNGpol,polcylinder}, charge transfer reactions, local optical absorption \cite{opticstobe} and charge carrier transport. As we discuss the low-frequency polaron dynamics in this paper within a \textit{simplified} theoretical framework, it will be particularly illustrated that the mobilities of the solvated carriers can be orders of magnitude lower than the intrinsic values. We note that the system we consider appears an interesting model realization in the general context of quantum particles interacting with a dissipative environment \cite{weiss99}. While we restrict our discussion to rigid SC nanostructures with good intrinsic electronic properties and favoring the formation of continuum adaibatic polarons, the generic aspects apply to other systems as well. Noteworthy, Basko and Conwell \cite{basko} emphasized a conceptually similar formation of hole polarons in DNA in aqueous electrolyte solutions and discussed the effect of water drag on their mobility \cite{CBdrag}.

The parametric scales for the spatial extent and the binding energy of the 3$d$ polaron are established by the 3$d$ Coulomb center problem where, for a carrier of charge $q$ and effective mass $m$ in the uniform medium of effective dielectric constant $\est$, the corresponding 3$d$ Bohr radius and Rydberg are \footnote{As we do not compare polarons and excitons in this paper, our definitions here are based on the electron mass rather than on the exciton reduced mass used in Refs.~\cite{YNGpol,polcylinder}.}
\begin{equation}\label{defaB}
\aB=\est \hbar^2 / m q^2, \ \ \ \ \Ry=q^2/2\est\aB.
\end{equation}
As a result of the confinement of charge carrier motion to the 1$d$ SC, the effects of Coulomb interactions are generally known \cite{haugbook} to get amplified, and, in fact, the binding energy $\bpol$ of the confined polaron can increase substantially in comparison with the 3$d$ binding \cite{YNGpol,polcylinder}. Our interest is in the strong confinement regime: $\aB \gg R$ ($R$ is the transverse radius of the 1$d$ structure), when the resulting longitudinal spatial extent of the polaron is much larger than $R$ (see Fig.~\ref{locmodes}(b)) and the self-localized charge carrier can be adequately described by the 1$d$ wave function $\wfx$, $x$ being the coordinate along the structure axis.

We consider the polaron to be formed due to the interaction of an excess carrier with the slow polarization component of the Debye solvent described by the frequency $\om$-dependent dielectric function
\begin{equation}\label{deb1}
\epom = \epsi + \frac{\epss - \epsi}{1- i\om\tad},
\end{equation}
for which the effective dielectric constant in Eq.~(\ref{defaB}) would be determined from the well-known \cite{polarons1,CTbook,nitzan} relationship
\begin{equation}\label{defest}
1/\est = 1/\epsi - 1/\epss,
\end{equation}
with $\epss \gg \epsi$ for typical solvent values of static $\epss$ and high-frequency $\epsi$ constants \cite{fawcett}. The characteristic time $\tad$ in Eq.~(\ref{deb1}) is
the Debye (transverse) relaxation time, for the problem at hand, however, the solvent response occurs at generally much shorter time scales corresponding to the longitudinal relaxation time \cite{nitzan,fawcett,frobook}
\begin{equation}\label{deftauL}
\tau_L=\frac{\epsi}{\epss}\,\tad.
\end{equation}
Typical solvents are charaterized by a wide range of $\tau_L$ ranging from fractions to tens of ps \cite{fawcett,YNGpol}. With the estimated polaron binding energies $\bpol$ on the order of 0.1 eV, the dynamic adiabaticity is ensured by $\hbar/\tau_L \ll \bpol$. For room temperatures $T$, on the other hand, one can also have $\kB T \ll \bpol$ satisfied, thus making thermal population of higher energy electronic states negligible.

The aim of this paper is to elucidate the response of such 1$d$ polarons to weak perturbations of low-frequencies $\om$: $\hbar \om \ll \bpol$, which do not cause transitions between different electronic levels. This, in particular, encompasses such basic phenomena as polaron drift caused by an applied field and polaron diffusion caused by the dielectric fluctuations of the surrounding solvent. We stress that our consideration is limited to a simplified theoretical model whose main purpose is to address elements of the essential physics. We will show how the underlying combined microscopic dynamics of the electronic charge density on the 1$d$ SC and the solvent results in point-particle-like Langevin dynamics of the polaron center. The derived mobility and diffusion coefficient, as should be expected \cite{weiss99}, satisfy the Einstein relation. In addition, we demonstrate the existence of local relaxation modes around the polaron whose relaxation times are somewhat longer than bare solvent's $\tau_L$. These modes involve polaron shape variations and presumably could manifest themselves in responses to time-dependent fields as well as in collisions.

\section{General adiabatic polaron relationships}

In the adiabatic picture, the excess charge carrier instantaneously responds to the  electrostatic potential $\potxt$ that is slowly changing in time by staying in the lowest eigenstate of the
\textit{stationary} Schr\"{o}dinger equation:
\begin{equation}\label{g1}
H \wfxt = E\wfxt, \ \ \
H=-\frac{\hbar^2}{2m}\frac{\partial^2}{\partial x^2} + q\potxt,
\end{equation}
where time $t$ thus enters only as a parameter.
The corresponding time-dependent 1$d$ charge density is
\begin{equation}\label{g2}
\denxt = q\, |\wfxt|^2,
\end{equation}
where $q$ is the charge of the carrier.

The electrostatic potential $\potxt$, in turn, can be represented as a sum of two contributions:
\begin{equation}\label{g2a}
\potxt = \potixt + \potsxt,
\end{equation}
where the first term represents the potential ``independent''
of the charge carrier itself, that is, the potential that would be there if the charge carrier was absent -- as exemplified by applied fields and dielectric fluctuations of the environment. The second term in (\ref{g2a}) is the part that is induced by the carrier charge density (\ref{g2}) in the past via the time-dependent medium response:
\begin{equation}\label{g2b}
\potsxt = \int_{0}^{\infty} d\tp \int d\xp \mathcal{G}(\xp,\tp) \rho (x-\xp,t-\tp).
\end{equation}
Equation (\ref{g2b}) assumes the translationally invariant response so that
$$
\potsko = \mathcal{G} (k,\om) \rho (k,\om).
$$
in Fourier space. The response function $\mathcal{G}$ is determined from the solutions of the corresponding 3$d$ electrostatic potential problem using the appropriate geometry of  elementary charge distributions and of the dielectric environment. Let us denote such a solution derived for a particular value $\epom$ of the medium dielectric function as $V(k,\epom)$.   One, of course, recognizes that $\mathcal{G}$ should \textit{not} include the instantaneous Coulomb interaction of the charge carrier with itself; in the context of the Debye solvent (\ref{deb1}), the medium for this case is characterized by $\epsi$. In these terms,
\begin{equation}\label{g2c}
\mathcal{G} (k,\om) = V(k,\epom) - V(k,\epsi),
\end{equation}
the relevant response is thus due to the slow (orientational) component of the dielectric polarization of the medium. Explicit examples of such calculations for the nanotube geometry with the account of tube's own polarizability can be found in Ref.~\cite{YNGpol}. When the dielectric polarization effects are well captured by the model of a uniform 3$d$ medium, which is a good approximation in the strong confinement regime \cite{YNGpol,tersscaling}, the response $\mathcal{G}$ is separable into a product of the spatial and temporal components (all bare Coulomb interactions are simply decreased by a factor of the inverse dielectric constant). We will write this as
\begin{equation}\label{g2d}
\mathcal{G}(x,t) = - \mathcal{A} (t) \,G (x) \ \ \ \mathrm{or} \ \ \ \mathcal{G} (k\om) = - \, \aom \,g (k).
\end{equation}
To gain more physical insight and obtain explicit simpler relationships, we will restrict our attention in this paper to this regime.
In accordance with Eq.~(\ref{g2c}), the temporal part for the Debye solvent (\ref{deb1}) can be described as
\begin{equation}\label{deb2}
\aom = \frac{1}{\epsi} - \frac{1}{\epom} =
\frac{\alo}{1-i\om\tau_L}, \ \ \  \alo=\frac{1}{\est}.
\end{equation}
The relaxation time $\tau_L$ that determines the pole of the \textit{inverse} dielectric function as well as the familiar combination (\ref{defest}) naturally arise here. As is well known \cite{frobook,nitzan,fawcett}, $\tad$ characterizes the dielectric relaxation in circumstances controlled by electrostatic potentials (electric fields $\vE$) while $\tau_L$ is appropriate in circumstances controlled by charge densities (electric inductions $\vD$). Evidently, the latter is the relevant situation here, and we will use $\tal\equiv\tau_L$ to simplify the notation in what follows. In the time domain, the response function is therefore
\begin{equation}\label{deb3}
\Alt = \int \frac{d\om}{2\pi}\, e^{-i\om t}\, \aom = \frac{\alo}{\tal}\, e^{-t/\tal} \,\Theta(t),
\end{equation}
where $\Theta$ is a step function.

The spatial part $G(x)$  of the response (\ref{g2d}) is determined by the specifics of the transverse charge distribution in our 1$d$ SC. For the illustrative calculations below, we will use the case of the nanotube geometry where (see, e.g., Refs.~\cite{ando,YNGpol})
\begin{equation}\label{tube1}
g (k) = 2 I_{0}(kR) K_{0} (kR),
\end{equation}
with $I_{0}$ and $K_{0}$ being the modified Bessel functions and $R$ the tube radius. Equation (\ref{tube1}) essentially corresponds to the Fourier transform of the Coulomb interaction between two charged rings of radius $R$.

Taking advantage of the particular simple form (\ref{deb3}), the temporal evolution of the potential $\potxt$, as per Eqs.~(\ref{g2a}) and (\ref{g2b}), can be conveniently converted into the differential (over time) equation:
\begin{eqnarray}
\frac{\partial \potxt}{\partial t}  +  \frac{\potxt}{\tal} & = &
\frac{\partial \potixt}{\partial t} + \frac{\potixt}{\tal} \hspace*{20mm} \nonumber \\
& - &
\frac{\alo}{\tal} \int d\xp G(x-\xp)\, \denxpt. \label{g5}
\end{eqnarray}

\section{The stationary solution and small perturbations around it}

One first discusses the dynamics in the absence of the external potential $\potixt$,
and the starting point, of course, is the consideration of the ground-state \textit{stationary} polaron
problem. In this basic case, the resulting charge density and potential are time-independent, we denote their spatial distributions as $\den (x)$ and $\pot (x)$, respectively. Their relationship is determined by the static medium response, or $\alpha (0)=\alo$ in Eqs.~(\ref{g2d}), (\ref{deb2}). That is also immediately seen from Eq.~(\ref{g5}):
\begin{equation}\label{g4}
\pot (x-\xo) = - \alo \int d\xp G(x-\xp) \, \den (\xp-\xo),
\end{equation}
where we explicitly emphasized the translational invariance by indicating that the stationary solutions can be centered around arbitrary points $\xo$. (The negative sign in the r.~h.~s. of Eq.~(\ref{g4}) and in other expressions below signifies that a positive charge density leads to a stabilizing negative potential, and \textit{vice versa}.) As $\den (x)$ is determined (\ref{g2}) by the ground-state charge carrier wave function $\psi_0 (x)$ obeying Eq.~(\ref{g1}), the latter wave function is then found from the self-consistent
non-linear Schr\"{o}dinger equation
$$
-\frac{\hbar^2}{2m}\frac{\partial^2 \wfx}{\partial x^2} \,-\, \frac{q^2}{\est}
\int \! d\xp G(x-\xp)\, |\wfxp|^2 \,\wfx = E\wfx
$$
as corresponding to its lowest eigenvalue $E=E_{0}$. This, of course, is the same equation one would derive from the optimization of the adiabatic system energy; it results in self-localized (polaronic) solutions as we discussed in Refs.~\cite{YNGpol,polcylinder}.

With the ground-state polaron solution in place, one can now consider the effects of
\textit{small} potential
perturbations $\dpotxt$: $\potxt = \pot (x) + \dpotxt$, in the stationary
Schr\"{o}dinger problem (\ref{g1}):
$
H=\overline{H} + V,
$ 
$$
\overline{H}=-\frac{\hbar^2}{2m}\frac{\partial^2}{\partial x^2} +
q\pot (x), \ \ \ \ V=q\dpotxt.
$$
The overall approach here is very similar to used in previous studies of small perturbations around 3$d$ \cite{Miyaki} and 1$d$ \cite{Melnik,SW78,holstein1} optical polarons. The standard lowest-order perturbation theory with respect to
$V$ then results in the perturbed electron density $\denxt = \den (x) + \dden (x,t)$ with
\begin{equation}\label{p2}
\dden (x,t)= - \int d\xp L(x,\xp) \, \dpotxpt,
\end{equation}
where the Hermitian kernel
$$
L(x,\xp)= q^2 \sum_{i > 0} \frac{\psi^*_0 (x)\psi_i (x) \psi^*_i
(\xp)\psi_0 (\xp) + \mathrm{c.c.}}{E_{i}-E_{0}}
$$
is built out of the solutions of the unperturbed problem
$$
\overline{H} \psi_i (x) = E_i \psi_i (x).
$$
The density perturbation (\ref{p2}) should self-consistently determine the potential perturbations (\ref{g2b}) at future times, that is:
\begin{equation}\label{p3}
\dpotxt= \int_{0}^{\infty}
d\tp\int d\xp \mathcal{G}(\xp,\tp) \, \dden (x-\xp,t-\tp).
\end{equation}
Together, Eqs.~(\ref{p2}) and (\ref{p3}) determine the eigenmodes of the un-driven dynamics around the stationary polaron in the case of a general response $\mathcal{G}$. In the separable case (\ref{g2d}), the problem of finding eigenmodes evidently reduces to a simpler generalized eigenvalue problem which we write in the form of
\begin{subequations}\label{p4}
\begin{eqnarray}
\denn (x) & = & - \int d\xp L(x,\xp)\, \potn (\xp), \label{p4a} \\
\potn (x) & = & - \lan\alo \int d\xp G(x-\xp)\,\denn (\xp).
\label{p4b}
\end{eqnarray}
\end{subequations}
In practical terms, the generalized problem (\ref{p4}) is easily converted to a normal eigenvalue problem and readily solved in Fourier (wave vector) space.
The eigenfunctions of this problem form a complete orthogonal set. To save space on notation, here we choose to normalize this set without extra coefficients as
\begin{equation}\label{p5}
-\int dx\, \denm (x)\, \potn (x) = \delta_{nm}
\end{equation}
(all functions can be chosen real). The eigenvalues $\lan \geq 1$ ($n \geq 0$), in turn, determine the eigenfrequencies $\omn$ from
\begin{equation}\label{p6}
\aomn = \lan \alo.
\end{equation}

\begin{figure}
\includegraphics[scale=0.7]{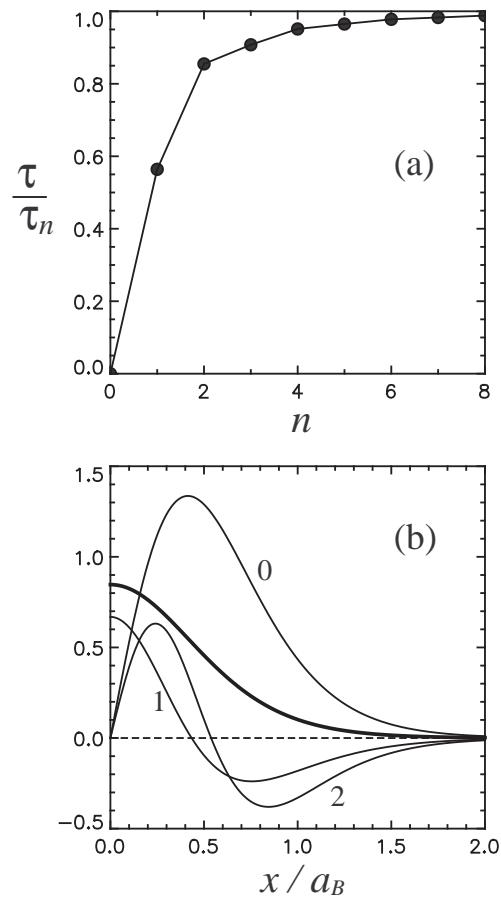}
\caption{\label{locmodes}Illustration of local relaxation modes (tubular geometry with $\aB/R=10$): (a) Relaxation times for the first $n=0$ to 8 eigenmodes; (b) Shapes of the spatial density variations $\denn(x)$ for the first 3 modes as indicated by values of the index $n$ next to thin lines. These lines are arbitrarily scaled here to be better displayed. The thick line corresponds to the scaled static polaron density $\aB \den (x) /q$.}
\end{figure}

Unlike the oscillating solutions in Refs.~\cite{Miyaki,Melnik,SW78,holstein1}, for the Debye solvent with its dissipative dynamics (\ref{deb2}), Eq.~(\ref{p6}) leads to a set of relaxation times $\taln$:
\begin{equation}\label{p7}
\omn = - \frac{i}{\taln}, \ \ \ \ \ \taln =
\frac{\tal}{1-1/\lan}.
\end{equation}
The temporal evolution of the eigenfunctions of the
dynamical problem would therefore be ``ordinarily'' decaying:
$$
\denn (x,t) \propto \denn (x) \, e^{-t/\taln}.
$$
Figure \ref{locmodes}(a) illustrates how these relaxation times $\taln$ converge to the ``bare'' value of $\tal$ upon the increase of the mode index $n$ - for this illustrative calculation we used the tubular case (\ref{tube1}) with the confinement parameter $\aB/R=10$.
One immediately notices, however, that the dynamics has a very different character for the mode with $\lambda_{0}=1$ corresponding to $\om_0 = 0$. This is the so-called zero-frequency translational mode whose existence is a general feature of quantum-field models with non-trivial spatial solutions and is a consequence of the
translational invariance \cite{raja82}. As discussed above, the static polaronic solutions are degenerate with respect to the position of the polaron center. An infinitesimal displacement along $x$ corresponds to the spatial derivative of the static solutions over $x$ (denoted below with symbol $^{\prime}$); one indeed easily verifies that
\begin{equation}\label{p8}
\deno (x) = \ga^{-1/2} \denp (x), \ \ \poto (x) = \ga^{-1/2} \potp
(x)
\end{equation}
satisfy Eqs.~(\ref{p4}) with $\lambda_{0}=1$. Here the normalization factor involves
\begin{equation}\label{p9}
\ga = \int dx |\denp (x) \potp (x)|.
\end{equation}
As an example, Fig.~\ref{locmodes}(b) shows spatial patterns corresponding to the first 3 eigenmodes in comparison with the static polaron shape in terms of the charge density. Of course, all density variations satisfy $\int dx\, \denn (x) =0$. Given the numerical nature of the solution, we have confirmed that the translational mode ($n=0$) precisely follows the spatial derivative of the static polaron shape. The next mode ($n=1$) can be qualitatively thought of as corresponding to the variation of the polaron width. The convergence pattern in panel (a) and the spatial variations displayed in panel (b) of Fig.~\ref{locmodes} are reminiscent of those discussed in connection with 1$d$ optical polarons \cite{Melnik,SW78,holstein1}.

As is well known \cite{raja82,holstein2}, in the
expansions of the arbitrary spatial patterns over the complete set of
eigenmodes, the zero-frequency mode should be excluded
in favor of the collective coordinate $\ksit$ of the
polaron center position (polaron centroid). Such expansions for the potential and density variations  then take the form of
\begin{subequations}\label{ex12}
\begin{eqnarray}
\potxt & = & \pot (\xrelt) + \sumn \ant \potn (\xrelt), \ \ \ \ \label{ex1}\\
\denxt & = & \den (\xrelt) + \sumn \ant \denn (\xrelt), \ \ \ \
\label{ex2}
\end{eqnarray}
\end{subequations}
where the the first terms correspond to the shapes of the static solutions that are centered around $\ksit$. The set of time-dependent coefficients
$\ant$ ($n \geq 1$) and the centroid coordinate $\ksit$ then
constitute complete co-ordinates of the system and can be conveniently used to describe its dynamics.

Using Eqs.~(\ref{ex12}) in Eq.~(\ref{g5}) for the potential evolution, one converts the latter into
\begin{equation}\label{m3}
-\dxi \left[\potp + \sumn \an \potnp  \right] + \sumn \left[
\dot{a}_n +\frac{\an}{\taln} \right]\potn = F (x,t),
\end{equation}
where the external perturbation part is
\begin{equation}\label{m3a}
F (x,t) = \frac{\partial \potixt}{\partial t} +
\frac{\potixt}{\tal}.
\end{equation}
As in Eqs.~(\ref{ex12}), all potential expansion functions in Eq.~(\ref{m3}) have the same argument $(\xrelt)$ .

\section{Equations of motion and centroid dynamics}

One derives the equations of motion for the set of coordinates $\{\ksit,\ant\}$ by projecting Eq.~(\ref{m3}) with each of $\denn (\xrelt)$ ($n \geq 0$) and utilizing the orthonormality relationships (\ref{p5}). Now note that we are interested in and have been developing our framework to study the lowest-order effects of small perturbations. Correspondingly, the system responses to such perturbations, the centroid velocity $\dxi (t)$ and shape deformation coefficients $\ant$, are small. Therefore, the second-order term in Eq.~(\ref{m3}) (containing products of $\dxi$ and $\an$) can safely be dropped. The result is then a set of fully decoupled equations of motion:
\begin{eqnarray}
\ga \, \dxi  & = & f, \label{m4a}\\
 \dot{a}_n +\an/\taln
 & = &  f_{n}, \ \ n\geq 1, \label{m4b}
\end{eqnarray}
where
\begin{equation}\label{m5}
f(\xi,t)= \int dx \denp (x) F(x+\xi,t)
\end{equation}
and
$$
f_{n} (\xi,t)= - \int dx \denn (x) F(x+\xi,t).
$$
Equations (\ref{m4a}) and (\ref{m4b}) evidently allow one to easily study various dynamic responses. Here we will concentrate on the centroid dynamics.

The external (perturbation) potential $\potixt$ can be represented as a sum of two contributions:
$$
\potixt=\potaxt+\potfxt,
$$
where $\potaxt$ is due to the applied field while $\potfxt$ is due to the ``bare'' fluctuations of the dielectric polarization (here, of course, only its slow, orientational, component is relevant). We will correspondingly use subscript indices $a$ and $f$.

Let us consider first the motion of the polaron under the action of a constant
applied potential $\potax$ that can slowly (on the scale of the
polaron size) change in space. From Eqs.~(\ref{m3a}) and (\ref{m5}) it immediately follows that
\begin{eqnarray}
f_a (\xi) & = & \int dx \denp (x) \frac{\pota (x+\xi)}{\tal}
 =  - \int dx \den (x) \frac{\partial \pota (x+\xi)}{\tal\,\partial \xi} \nonumber \\
& \simeq & -\left( \frac{\partial \pota (\xi)}{\tal\,\partial \xi}
\right) \int dx \den (x) = \frac{q \fE (\xi)}{\tal}, \label{c3}
\end{eqnarray}
where $\fE$ is the electric field acting on the polaron.

On the other hand, the fluctuating contribution $f_f (\xi,t)$ is a random process whose correlations have to be determined from the correlations of $\potfxt$ subsequently using  definitions (\ref{m3a}) and (\ref{m5}). The bare classical ($\kB T \gg \hbar/\tal$) fluctuations  are easily found by using the fluctuation-dissipation theorem for the specified 1$d$ geometry, the result is
\begin{equation}\label{c6}
\avg{\potf (x,t) \potf (\xp,\tp)} = \alo \kB T \, G(x-\xp) \,
e^{-|t-\tp|/\tal}.
\end{equation}
With Eq.~(\ref{m3a}), this translates \textit{exactly} into
\begin{equation}\label{c7}
\avg{F_f (x,t) F_f (\xp,\tp)} = \frac{2 \alo \kB T}{\tal} \, G(x-\xp)
\, \delta (t-\tp).
\end{equation}
Given the time-uncorrelated character of Eq.~(\ref{c7}),
the relevant correlations of $f_f$ are those for the same
centroid coordinate:
\begin{equation}\label{c8a}
\avg{f_f (\xi,t)f_f (\xi,\tp)} = \frac{2 \ga \kB T}{\tal}\,\, \delta (t-\tp).
\end{equation}
To derive Eq.~(\ref{c8a}) from Eqs.~(\ref{m5}) and (\ref{c7}), we have also used Eq.~(\ref{g4}) in the following:
$$
\int dx\,d\xp \denp (x) \denp (\xp) \, G(x-\xp) =
 -\int dx \frac{\denp (x) \potp (x)}{\alo} = \frac{\ga}{\alo}.
$$

Introducing the random velocities $\eta (t)= f_f(t)/\ga$, the equation of motion (\ref{m4a}) for the polaron centroid under the action of a constant field can be conveniently rewritten as
\begin{equation}\label{c9}
\dxi = \mu \,q \fE + \eta (t),
\end{equation}
where the mobility
\begin{equation}\label{c10}
\mu = 1/\ga\,\tal 
\end{equation}
is defined with respect to the dynamical force $q\mathcal{E}$ while
the random process $\eta (t)$, as per Eq.~(\ref{c8a}), satisfies
\begin{equation}\label{c11}
\avg{\eta(t)\eta(\tp)}=2D \delta(t-\tp)
\end{equation}
with the diffusion coefficient
\begin{equation}\label{c12}
D = \kB T / \ga\,\tal. 
\end{equation}
Equation (\ref{c9})  is the standard overdamped
Langevin equation \cite{nitzan} for a point particle as consistent with our assumption of the field having practically no spatial variation on the scale of the polaron size. We emphasize that ingredients of Eq.~(\ref{c9}) have been \textit{derived} from the underlying microscopic dynamics. Our
mobility (\ref{c10}) and diffusion coefficient (\ref{c12}) manifestly obey
the Einstein relation
$$
D = \mu \kB T,
$$
thus ensuring that the macroscopic dynamics of the polaron in the
steady state leads to the Boltzmann distribution. We note that expression (\ref{c10}) is similar in form to the one obtained from energy conservation considerations by Basko and Conwell \cite{basko,CBdrag} but with the important quantitative distinction that our $\tal=\tau_L$ is the longitudinal relaxation time rather than $\tad$ they used (the difference that can easily exceed an order of magnitude).

Our discussion of the constant applied field is straightforwardly generalized to time-dependent fields. As Eq.~(\ref{m3a}) suggests, for fields varying as $\fE (t) = \fE_{0}\exp(-i\omega t)$, the first term in Eq.~(\ref{c9}) would get modified, with $1/\tal$ being replaced by $-i\omega + 1/\tal$. In addition, time-dependent fields would be exciting polaron shape deformations (\ref{m4b}) as we will discuss elsewhere.

\begin{figure}
\includegraphics[scale=0.7]{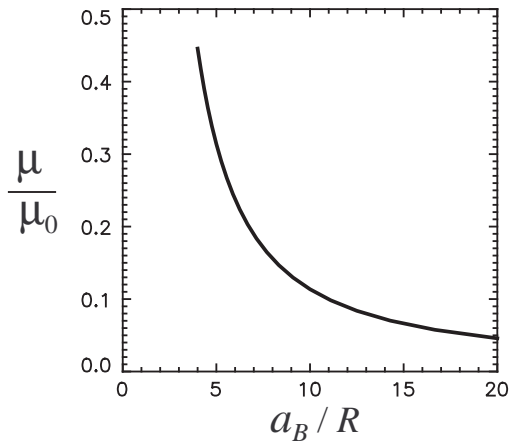}
\caption{\label{mobility}The polaron mobility in the constant applied field as a function of the confinement parameter $\aB/R$, here the mobility is measured in units of $\mu_0=\aB^2/\tau_L\Ry$ which itself can strongly depend on system parameters.}
\end{figure}

The calculation of the mobility (\ref{c10}), within the studied model, is thus reduced to the evaluation of Eq.~(\ref{p9}) based on the ground-state of the polaron. Figure \ref{mobility} displays results of such a calculation for the tubular interaction (\ref{tube1}) in the scaled form that separates the specific dependence of the mobility on the confinement parameter $\aB/R$. Within the shown region of its variation, the behavior exhibited in Fig.~\ref{mobility} may be \textit{approximated} by a power-law:
\begin{equation}\label{sc1}
(\mu/\mu_0) \propto (\aB/R)^{-1.4}.
\end{equation}
For the overall dependence of the polaron mobility on system parameters, one should recall that the reference value of
\begin{equation}\label{sc2}
\mu_0=\aB^2/\tau_L\Ry
\end{equation}
is itself strongly parameter-dependent as per Eqs.~(\ref{defaB}) and (\ref{deftauL}).

Using Eqs.~(\ref{sc1}) and (\ref{sc2}), one can establish how the mobility scales when some of system parameters vary. The effect of the radius $R$ on the mobility is already explicit in Eq.~(\ref{sc1}). For the dependence on the effective mass $m$, one finds an approximate scaling law
\begin{equation}\label{sc3}
\mu \propto m^{-1.6}.
\end{equation}
The dielectric properties of the solvent are also important here: if one were to separate the effect of $\tau_L$ on $\mu$ the remaining factor for the mobility would feature the dependence on $\est$ as
\begin{equation}\label{sc4}
\tau_L \cdot \mu \propto (\est)^{2.6}.
\end{equation}
With typical $\epss \gg \epsi$, $\est \simeq \epsi$. To appreciate the effect on the mobility itself, however, one is reminded that the longitudinal time  $\tau_L$ (\ref{deftauL}) is also proportional to $\epsi$. Equations (\ref{sc3}) and (\ref{sc4}) illustrate a quite understandable general trend: the more strongly bound the polaron is, then the lower its mobility would be.

\section{Discussion}

One-dimensional semiconductor nanostructures in immediate contact with polar liquids is an interesting class of systems of relevance to many applications, particularly those involving redox reactions \cite{kamat07,SCelectrodes}. Using simplified theoretical models, we have demonstrated \cite{YNGpol,polcylinder} that the solvation of excess charge carriers due to the solvent polarization can cause carriers on 1$d$ SCs to self-localize. A combination of several factors, small intrinsic effective masses of the carriers, enhanced Coulomb effects in 1$d$ (small-diameter structures) and the sluggish nature of the orientational solvent polarization,  creates conditions in these systems that may be favorable for the resulting formation of 1$d$ continuum adiabatic polarons. The physical properties of the polarons are quite different from those of the band states. The energetic significance of the polaronic effect is emphasized by making comparisons with much more studied Wannier-Mott excitons in 1$d$ SCs. When compared theoretically on an equal footing, we have found \cite{YNGpol,polcylinder} that the polaron binding energy, $\bpol$, can be a substantial fraction, roughly one-third for charge-conjugation-symmetric SCc, of the exciton binding energy, $\bexc$, in the media with comparable $\epsi$.

Let us take a widely known example of single-wall carbon nanotubes (SWCNTs). The importance of the excitonic effects in the optics
of semiconducting SWCNTs is well established now, with binding
energies $\bexc$ experimentally measured in some tubes to be in
the range of $0.4 - 0.6$ eV \cite{optrescnt,bachilo_bind,krauss1}. That would suggest that polaron binding energies can reach values of 0.1 eV and more, as it also follows from numerical estimates, reaffirming the potential importance of polarons. It would be pertinent to note that redox chemistry of CNTs has been deemed an ``emerging field of nanoscience'' \cite{chirsel} and solvatochromic effects in CNTs are being intensely researched \cite{CNTsolvchrom}; interestingly, even the first mapping of luminescence versus absorption spectra of individual SWCNTs has been achieved in aqueous suspensions. \cite{micelle1,micelle2}

In an effort to establish experimentally testable signatures of polaron formation, in this paper we have studied the low-frequency dynamics of 1$d$ large adiabatic polarons. We have shown how the combined microscopic dynamics of the electronic charge density on the 1$d$ SC and the solvent results in macroscopic Langevin dynamics of the polaron centroid. The mobility (\ref{c10}) and diffusion coefficient (\ref{c12}) are thus derivable from the stationary polaron parameters and the solvent dynamic relaxation. We also demonstrated that new local dielectric relaxation modes develop in the presence of the polaron. These modes involve polaron shape deformations and may play a role in time-dependent responses and in polaron collision events.

It is instructive to make some numerical estimates based on the obtained results. If, for instance, one chooses a set of reasonable representative numerical parameters: $m=0.05 m_e$ ($m_e$ being the free electron mass), $\est=3$ and $\tau_L=1$ ps, then, by Eq.~(\ref{defaB}), the resulting  $\aB \simeq 32$ {\AA}, $\Ry \simeq 76$ meV, and, by Eq.~(\ref{sc2}), $q \cdot \mu_0 \simeq 1.3$ cm$^2$/V$\cdot$s. Using the data of Fig.~\ref{mobility} for $\aB/R=6$ as an example of a small-diameter structure, this value of $\mu_0$ would then translate into $q\cdot \mu \simeq 0.3$ cm$^2$/V$\cdot$s. At T=300 K this yields the value of the diffusion coefficient $D \simeq 8 \times 10^{-3}$ cm$^2$/s. Of course, with a wide range of variability of the relevant system parameters (suffice to mention values of $\tau_L$), these tentatively estimated values can change substantially for various systems. Still, these numbers strongly indicate that the mobility of the solvated carriers can be orders of magnitudes smaller than the intrinsic carrier mobilities of 10$^3$-10$^5$ cm$^2$/V$\cdot$s discussed for SC wires and tubes \cite{NWelectronics,NTelectronics}.

We conclude by reiterating that the illustrative calculations presented in this paper are based on simplified generic models whose main role is to elucidate the essential physics. While we do not expect substantial changes in the results valid for small-diameter nanostructures (the strong-confinement regime, when most dielectric effects are due to the surrounding medium \cite{YNGpol,polcylinder,tersscaling}), more accurate calculations for specific systems, including a finite size of dielectric cavities, SC's band structure and polarizability, may be required for detailed comparison with prospective experimental data.\\

We are grateful to V.~M.~Agranovich and E.~M.~Conwell for useful
comments. This study was supported by the Collaborative
U.~T.~Dallas -- SPRING Research and Nanotechnology Transfer
Program.

\bibliography{fluctuations}

\end{document}